\documentclass[%
 reprint,
 superscriptaddress,
 amsmath,amssymb,
 aps,
prd,
]{revtex4-2}

\usepackage{float}
\usepackage{bbold}
\usepackage{braket}
\usepackage{amssymb,amsmath}

\usepackage{bbm}

\usepackage{amsfonts}
\usepackage{subfigure}
\usepackage{array}
\usepackage{enumerate}

\usepackage{graphicx}
\usepackage{dcolumn}
\usepackage{bm}
\usepackage[dvipsnames]{xcolor}
\usepackage[colorlinks, citecolor=blue,anchorcolor=red,menucolor=red,linkcolor=red,filecolor=red,runcolor=red,urlcolor=blue,frenchlinks=red]{hyperref}
\usepackage[nameinlink]{cleveref}
\crefformat{equation}{Eq.~(#2#1#3)}
\crefformat{figure}{Fig.~#2#1#3}
\crefformat{section}{Sec.~#2#1#3}

\usepackage{ulem}

\graphicspath{{./figures/}{./}}

\begin{document}


\title{Magnetic catalysis and diamagnetism from pion fluctuations}

\author{Jie Mei}
\email{meijie22@mails.ucas.ac.cn}
\affiliation{School of Nuclear Science and Technology, University of Chinese Academy of Sciences, Beijing, 100049,  P.R. China}
\affiliation{School of Physical Sciences, University of Chinese Academy of Sciences, Beijing 100049, P.R. China}
\affiliation{Institute of High Energy Physics, Chinese Academy of Sciences,Beijing, 100049,P.R. China}

\author{Rui Wen}
\affiliation{School of Nuclear Science and Technology, University of Chinese Academy of Sciences, Beijing, 100049,  P.R. China}

\author{Shijun Mao}

\affiliation{School of Physics, Xi'an Jiaotong University, Xi'an, Shaanxi 710049, P.R. China}

\author{Mei Huang}
\affiliation{School of Nuclear Science and Technology, University of Chinese Academy of Sciences, Beijing, 100049,  P.R. China}

\author{Kun Xu}
\email{xukun21@ucas.ac.cn}
\affiliation{School of Nuclear Science and Technology, University of Chinese Academy of Sciences, Beijing, 100049,  P.R. China}

\begin{abstract}
In the framework of Nambu--Jona-Lasinio model beyond mean field approximation, the effects of pion fluctuations on (inverse) magnetic catalysis and magnetic susceptibility are studied. The negative magnetic susceptibility at low temperature is observed when contributions from both neutral and charged pions are taken into account. In weak field approximation, it is observed that at finite temperature, the magnetic inhibition effect in the chiral limit, resulting from the difference between the transverse and longitudinal velocities of neutral pions, converts to weak magnetic catalysis when considering a non-zero current quark mass. Moreover, the magnetic catalysis is amplified by the charged pions. Therefore, no inverse magnetic catalysis is observed when considering pion fluctuations.
\end{abstract}

\maketitle


\section{INTRODUCTION}
\label{sec:level1}

The investigation of the response of quark matter to a uniform magnetic field background has been a hot topic for the last decade, see reviews \cite{Preis:2012fh, Gatto:2012sp, DElia:2012ems, Miransky:2015ava, Andersen:2014xxa}. In the experimental aspect, a strong but transient magnetic field can be generated in the initial stage of heavy ion collisions (HIC), which was believed to be the strongest magnetic field ever created, with the strength of $eB\sim10^{18-20}\ \text{Gau\ss}$ and life time of $10^{-22}$ second \cite{Kharzeev:2007jp, Gursoy:2014aka, delaIncera:2010wn, Duncan:1992hi,Li:2023tsf}. In the theoretical aspect, the interplay between a magnetic field and quantum chromodynamics (QCD) can lead to various novel behaviors and can be used as a probe to investigate the inner structure of quark matter.

The surge of the interest in studying the influence of magnetic field to a equilibrium system starts from two abnormal results from a lattice QCD group's ab-initio calculation, (i) although the vacuum quark mass is enhanced by the magnetic field, the chiral critical temperature decrease with the increasing of it, which is called inverse magnetic catalysis (IMC), (ii) the magnetic susceptibility is negative at low temperature while positive at high temperature \cite{Bali:2012edd,Bornyakov:lattice,DElia:lattice,Endrodi:lattice1,Endrodi:lattice2,Bali:2020bcn}. These results disagree with most of the effective model predictions at that time, for instance, the standard Nambu-Jona-Lasinio (NJL) model and linear-$\sigma$ model with quark (Quark-Meson model)  under mean-field approximation\cite{Klevansky:1989vi,Gusynin:1995nb}. 

Numerous studies have been conducted to elucidate the IMC and diamagnetic effects. These investigations have explored various mechanisms, including magnetic inhibition resulting from fluctuations of neutral pions~\cite{Fukushima:2012kc}, chirality imbalance stemming from sphaleron transitions or instanton-anti-instanton pairings~\cite{Chao:2013qpa,Yu:2014sla}, and the influence of the running coupling constant in the presence of a magnetic field~\cite{Ferrer:2014qka}. 
Some groups tried to include the anomalous magnetic moment effect \cite{Fayazbakhsh:2014mca,Mei:2020jzn,Xu:2020yag,Lin:2022ied,Kawaguchi:2022dbq,Tavares:2023oln,Chaudhuri:2023djv,Islam:2023zyo} or the effect of tensor channel \cite{Lin:2022ied} in the NJL model to reproduce the IMC or diamagnetism. By considering the running coupling with $eB$-dependence fitted by lattice QCD data, the IMC result can be successfully reproduced in \cite{Ferreira:2014kpa, Liu:2016vuw, Avancini:2016fgq, Ayala:2016bbi, Farias:2016gmy, Chen:2021gop, Sheng:2021evj, Mao:2022nfs}. Hadron resonance gas (HRG) model, where the hadrons are assumed as point-like particles with no interaction in between, can also reproduce diamagnetic result at low temperature region \cite{Kamikado:2014bua,Sahoo:2023vkw}. Functional continuum field approaches, such as the functional renormalization group (FRG), Dyson-Schwinger equations (DSE), and holographic QCD models also have made great efforts on both the effective models \cite{Wen:2023qcz, Li:2019nzj, Kamikado:2013pya, Kamikado:2014bua, Fukushima:2012xw, Fu:2017vvg,Bohra:2019ebj,Bohra:2020qom,Wen:2024hgu} and QCD theory \cite{Mueller:2014tea, Mueller:2015fka, Braun:2014fua}. In general, the understanding of the IMC and diamagnetic effects remains an open question.

Most of the work in the NJL model was done based on the mean-field approximation which only considers the lowest-order in  $1/N_c$ expansion \cite{Klevansky:1992qe}. In next-to-leading order of this expansion, the feedback effect from mesons is taken into consideration \cite{Quack:1993ie,Zhuang:1994dw}. Generally, mean field approximation for quark together with random phase approximation for meson works well to describe the thermodynamic properties of QCD matter in absence of magnetic field. For the puzzle of IMC and diamagnetism, the feedback from mesons can be part of the solution, given that they are influenced by magnetic field in both direct (for charged mesons) and indirect (for neutral mesons) manners \cite{Mao:2018dqe,Li:2021swv,Mei:2022dkd,Mao:2022nfs,Liu:2018zag,Coppola:2019uyr,Avancini:2021pmi,Li:2023rsy}. In references \cite{Fukushima:2012kc,Mao:2016fha,Mao:2017tcf}, the feedback effect from $\pi_0$ with a physical propagating velocity is included in the chiral limit, giving an IMC result. Besides, in the (Polyakov-loop extended) quark-meson model, the meson fluctuations, especially the light pion contributions, lead to the diamagnetism \cite{Li:2019nzj, Kamikado:2014bua}. In our calculation, we investigated beyond mean field by including the effect from both $\pi_0$ and $\pi^{\pm}$ with and without finite propagating velocities, and consider a physical situation where chiral symmetry is explicitly broken, to see their role in (inverse) magnetic catalysis and magnetic susceptibility.

This paper is arranged as follows: In \cref{sec:level2} we introduce the calculating procedure in beyond mean-field NJL model in the manner of weak-field expansion. In \cref{sec:results}, we give the numerical results for $eB$-dependence of critical temperature $T_c$ and magnetic susceptibility followed by a summary and discussion in \cref{conclusion}.

\section{The NJL model}
\label{sec:level2}

We start with the Lagrangian of SU(2) NJL model in the presence of a uniform magnetic field,
\begin{align}
	\mathcal{L}=\bar{\psi}\left(i\gamma^{\mu}D_{\mu}-\hat{m}\right)\psi+G\left[\left(\bar{\psi}\psi\right)^2+\left(\bar{\psi}i\gamma_5\vec{\tau}\psi\right)^2\right].
\end{align}
Here, $\psi$ is two-flavor quark field $\psi=(u,d)^T$, $\hat{m}$ is the quark current mass matrix $\hat{m}=\text{diag}(m_u,m_d)$ with $m_u=m_d=m_0$, which explicitly breaks the chiral symmetry, and $\tau_i$ is $i$-th component of Pauli matrices. The covariant derivative $D_\mu=\partial_\mu-i Q A_\mu$ coupling quarks with electric charge $Q=\text{diag}(q_u,q_d)=\text{diag}(2/3 e,-1/3 e)$ to a gauge field ${\bf B}=\nabla\times{\bf A}$. In this article, we choose the Landau gauge, where the potential $A_{\mu}=(0,0,Bx_1,0)$.

In Schwinger scheme, the translational invariance for a charged particle is broken by the magnetic field, deviding the quark propagator into two parts, namely, the Schwinger phase part and the Fourior transformation of translational invariant part.
\begin{align}\label{propagator_wf}
	S_f\left(x,y\right)=e^{i\Phi_f(x_{\perp},y_{\perp})}\int\!\!\frac{d^4 k}{(2\pi)^4} e^{-i k (x-y)} \tilde{S}_f\left(k_{\perp},k_{\parallel}\right).
\end{align}
Here, the parallel and perpendicular components of the four-dimensional coordinate and momentum are written as $x_{\perp}=(x_1,x_2)$, $k_{\perp}=(k_1,k_2)$ and $k_{\parallel}=(k_0,k_3)$. The quark Schwinger phase in Landau gauge reads
\begin{eqnarray}
	\Phi_f(x_{\perp},y_{\perp})=\frac{q_f B}{2} (x_1+y_1)(x_2-y_2),
\end{eqnarray}
which is gauge-dependent. The detailed forms of Schwinger phase in different gauges can be found in \cite{Miransky:2015ava,GomezDumm:2023owj}. To include the effect of Schwinger phase in weak field expansion, one can shift the the zeroth order of it. In \cref{sec:results}, we will explain that the missing of Schwinger phase does not affect the qualitative conclusion of this work, and we neglect Schwinger phase in the following. 

In the weak magnetic field limit, the quark propagator in momentum space can be expanded in the power of $(q_f B)$, and reads
\begin{align}
	&i\tilde{S}_f(k_{\perp},k_{\parallel})\nonumber \\
	&=i\frac{m_q+k\!\!\!/}{k^2-m_q^2}-\left(q_f B\right) \frac{\gamma_1\gamma_2\left(m_q+k\!\!\!/_{\parallel}\right)}{(k^2-m_q^2)^2}\nonumber\\
	&\ \ -2i\left(q_f B\right)^2\frac{k_{\perp}^2 (m_q+k\!\!\!/_{\parallel})+k\!\!\!/_{\perp} (m_q^2-k_{\parallel}^2)}{(k^2-m_q^2)^4}+\mathcal{O}(q_fB)^3 \nonumber\\
	&=i\tilde{S}^{(0)}+(q_f B)i\tilde{S}^{(1)}+(q_f B)^2i\tilde{S}^{(2)}+\mathcal{O}(q_fB)^3
\end{align}
where the contributions of transverse and parallel momentum are considered separately. Here, $m_q$ is the constituent quark mass, which is the same value for both $u$ and $d$ flavor. We use the notation,
\begin{eqnarray}
	(a\cdot b)_{\parallel}=a^0 b^0-a^3 b^3,\nonumber\\
	(a\cdot b)_{\perp}=a^1 b^1+a^2 b^2,
\end{eqnarray}
hence  $k^2=k_{\parallel}^2-k_{\perp}^2$. 

\subsection{Mean field approximation}

In mean field approximation, the thermodynamic potential of the system at finite temperature and under external magnetic field $B$ takes the form of:
\begin{eqnarray}
	\Omega_{\text{MF}}(T,B)=\frac{(m_q-m_0)^2}{4G}+\Omega_q(T,B),
\end{eqnarray}
with the contribution from quarks:
\begin{eqnarray}
	\Omega_{q}(T,B)=\text{Tr}_{\{c,f,s,x\}}\ln\left(\frac{1}{T}S^{-1}(x,x)\right).
\end{eqnarray}
The trace operation is carried out over color (c), flavor (f), spinor (s) degrees of freedom, as well as over the four-dimensional coordinate ($x$). $S^{-1}(x,x)$ is the inverse of quark propagator in coordinate space.

To determine the effective quark mass, we have to find the ground state by locating the global minimum of the thermodynamic potential
\begin{eqnarray}
	\frac{\partial\Omega_{\text{MF}}(T,B)}{\partial m_q}=0,\nonumber\\
	\frac{\partial^2\Omega_{\text{MF}}(T,B)}{\partial m_q^2}\geq 0,
\end{eqnarray}
which leads to the gap equation
\begin{align}\label{gap_mf}
	m_0=m_q-2G\text{Tr}_{\{c,f,s,k\}}[i\tilde{S}(k)].
\end{align}

Inserting \cref{propagator_wf} into \cref{gap_mf} and apply the summation over Matsubara frequency, we can obtain the weak-field expansion version of gap equation:
\begin{align}
    m_q\Biggl(1-2GN_c\sum_{f}\biggl(I_1^{(0)}+(q_f B)^2I_1^{(2)}+\mathcal{O}(q_f B)^4\biggr)\Biggr)=m_0,
\end{align}
with
\begin{align}
    I_1^{(0)}&=-4\int \frac{d^3 k}{(2\pi)^3} \mathcal{F}_{(1)}(m_q^2), \\
    I_1^{(2)}&=8\int \frac{d^3 k}{(2\pi)^3} k_\perp^2 \mathcal{F}_{(4)}(m_q^2).
\end{align}
There is no contribution from odd terms of $I_1^{(n)}$ after performing the trace in spinor space, which agrees with symmetry analysis. The contributions from higher orders of magnetic field are neglected. Here we define the fermionic threshold functions
\begin{align}
    \mathcal{F}_{(n)}(m_q^2)\equiv T \sum_{n_T} \frac{1}{(k^2-m_q^2)^n},
\end{align}
then it is straightforward to obtain:
\begin{align}
    \mathcal{F}_{(1)}&= \frac{1}{2 E_q}(1-2n_f(m_q^2)), \\
    \mathcal{F}_{(n+1)}&=\frac{1}{n} \frac{\partial}{\partial m_q^2}\mathcal{F}_{(n)}.
\end{align}
Here, $n_f(m_q^2)$ is fermionic distribution functions and quark energy dispersion relation $E_q=\sqrt{\mathbf {k}^2+m_q^2}=\sqrt{k_3^2+k_{\perp}^2+m_q^2}$.


 Correspondingly, the quark part of thermodynamic potential can also be rewritten in this weak-field expansion manner:
\begin{align}
	\Omega_{q}&=\Omega_q^{(0)}+\sum_{f}(q_fB)^2\Omega_q^{(2)}+\mathcal{O}(q_fB)^4, \\
	\Omega_q^{(0)}&=-2N_cN_f\int\!\!\frac{d^3 k}{(2\pi)^3}\left[E_q+2T\ln\left(1+e^{\frac{-E_q}{T}}\right)\right],\\
	\Omega_q^{(2)}&=-\frac{4N_c}{3}\int\frac{d^3 k}{(2\pi)^3}k_{\perp}^2\mathcal{F}_{(3)}(m_q^2). 
\end{align}

\subsection{Meson section}

In the NJL model, mesons are treated as quantum fluctuations above the mean field. Through the random phase approximation (RPA) method, the meson propagator can be expressed in terms of the irreducible polarization function or quark bubble,
\begin{align}\label{pimm}
	\Pi_M(q^2)=i\int\!\!\frac{d^4 k}{(2\pi)^4}\text{Tr}_{\{c,f,s\}}\left[\Gamma_{M}^* \tilde{S}(k)\Gamma_{M} \tilde{S}(k-q)\right],
\end{align}
with the meson vertex
\begin{equation}
	\Gamma_M=\left\{
	\begin{array}{l}
		1,\\
		i\gamma_5\tau_+,\\
		i\gamma_5\tau_-,\\
		i\gamma_5\tau_3,
	\end{array}	\right. 
    \begin{array}{l}
    	M=\sigma\\
    	M=\pi_+\\
    	M=\pi_-\\
    	M=\pi_0,
    \end{array},
\end{equation}
where $\tau_{\pm}$ is determined by $\tau_{\pm}=(\tau_1\pm i \tau_2)/\sqrt{2}$. By inserting \Cref{propagator_wf} into \Cref{pimm}, we can get the polarization function for $\pi_0$ in weak field approximation
\begin{eqnarray}
	\Pi_{\pi_0}(q^2)=N_c\sum_{f}\left[\Pi^{00}+(q_f B)^2\left(2\Pi^{20}+\Pi^{11}\right)\right],
	\label{piwf}
\end{eqnarray}
with definition
\begin{eqnarray}
	\Pi^{lm}(q^2)=i\int\!\!\frac{d^4 k}{(2\pi)^4}\text{Tr}_{\{s\}}\!\left[i\gamma_5 \tilde{S}^{(l)}(k)i\gamma_5 \tilde{S}^{(m)}(k-q)\right],\ \ 
	\label{piij}
\end{eqnarray}
After performing the trace in spinor space, the contribution of $\Pi^{10}$ and $\Pi^{01}$ vanishes. The calculation for Eq.~(\ref{piij}) is straightforward but tedious. For the polarization function for charged pions, it has the similar form as the neutral pion case,
\begin{align}
	&\Pi_{\pi_{\pm}}(q^2)\nonumber\\
	&=2N_c\left[\Pi^{00}+(q_uB)(q_dB)\Pi^{11}+\sum_{f}(q_f B)^2\Pi^{20}\right].
	\label{picwf}
\end{align}
Again, the contribution of $\Pi^{01}$ vanishes after carrying out the trace in spinor space.

Via taking the bubble summation in random phase approximation, the effective propagator for a meson $M$ can be constructed by
\begin{eqnarray}
	U_M=\frac{2G}{1-2G\Pi_M},
\end{eqnarray}
and the pole mass $m_{pole}$ (static solution, setting $q_1=q_2=q_3=0$) and the screening masses $m_{scr,i}$ in $q^i$ direction (setting $q_0=0$, and $q_j=0$ for $j \neq i$) can be solved by following equations, correspondingly,
\begin{eqnarray}\label{eq:poleeq}
	1-2G\Pi_M(q_0^2=m_{pole}^2,0)=0
\end{eqnarray}
and
\begin{eqnarray}
	1-2G\Pi_M(0,q_i^2=-m_{scr,i}^2)=0.
\end{eqnarray}
 The detailed calculation for the polarization function in weak-field expansion are listed in \cref{appendixA}. It should be noted that when meson pole (screening) mass exceeds the threshold of the mass sum of its constituent quarks, the meson undergoes a Mott transition and a finite width ($m_M\to m_M-i\frac{\Gamma}{2}$) should be taken into consideration. The meson mass and its width can be determined by the corresponding complex equations. In this paper, we work in the temperature region where pions are still bound state particles.

\subsection{Beyond mean-field approximation}

We go beyond the mean-field approximation by using the $1/N_c$ expansion \cite{Quack:1993ie, Zhuang:1994dw}. Another self-consistent beyond mean-field method in the NJL model is the FRG approach \cite{Fu:2022uow, Fu:2024ysj}. By including the next-to-leading order of $1/N_c$ expansion, meson degrees of freedom are self-consistently introduced.  The thermodynamic potential of the quark-meson plasma \cite{Zhuang:1994dw} can be rewritten as
\begin{eqnarray}
\label{eq:bmf_potential}
	\Omega=\frac{(m_q-m_0)^2}{4G}+\Omega_q+\sum_M\Omega_M.
\end{eqnarray}
In pole approximation, the meson contribution in thermodynamic potential with vanishing magnetic field takes the form,
\begin{eqnarray}
	\Omega_{\pi}=\int\!\!\frac{d^3k}{(2\pi)^3}\left[\frac{E_{\pi}}{2}+T\ln \left(1-e^{-E_{\pi}/T}\right)\right],
\end{eqnarray}
with meson energy dispersion relation
\begin{align}
	E_{\pi}=\sqrt{m_{\pi,pole}^2+(\mathbf{v}\cdot \mathbf {k})^2}.
	\label{eq:mesonE}
\end{align}
As shown in \cite{Shuryak:1990ie,Pisarski:1996mt,Fayazbakhsh:2012vr,Fayazbakhsh:2013cha,Sheng:2020hge}, $\bf{v}$ in \cref{eq:mesonE} represents the propagating velocity for the corresponding meson, whose $i$-th component has the form of \cite{Pisarski:1996mt,Fayazbakhsh:2012vr,Fayazbakhsh:2013cha,Sheng:2020hge}
\begin{align}
v_i=\frac{m_{M,pole}}{m_{M,scr,i}}.
\label{eq:v}
\end{align} 
For neutral pion, when we consider its behaviors in the hadron level, it has no interaction with magnetic field at all. However, in a model in quark level like NJL model, it's magnetic field-sensitive constituent quark will contribute to the variation of $\pi_0$ mass and the splitting of its propagating velocity in parallel and perpendicular directions, leading to the new form of energy dispersion relation
\begin{align}
E_{\pi_0}=\sqrt{m_{pole,\pi_0}^2+v_{\perp}^2k_{\perp}^2+v_{\parallel}^2k_{3}^2},\label{eq:NMeson}
\end{align}
which is believed to be responsible for the IMC effect in the chiral limit, as in \cite{Fukushima:2012kc,Mao:2016fha,Mao:2017tcf}. In absence of a uniform magnetic field, the neutral pion $\pi_0$ and charged pions $\pi_{\pm}$ are isospin triplet that share the same form of thermodynamic potential. When a magnetic field is turned on, the charged pions obtain extra masses from magnetic field, and decouple from neutral pion, which means we should consider their effect seperately. For charged pions, the magnetic field not only alter their pole masses and propagating velocities, but also change the momentum in perpendicular directions from thermodynamic potential and energy dispersion relation into discrete Landau levels,
\begin{align}
	&\Omega_{\pi_{\pm}}=\sum_{n=0}^{\infty}\frac{|eB|}{2\pi}\int_{-\infty}^{\infty}\!\!\frac{dk_3}{2\pi}\left[\frac{E_{\pi_{\pm}}}{2}+T\ln \left(1-e^{-E_{\pi_{\pm}}/T}\right)\right],\\
	&E_{\pi_{\pm}}=\sqrt{m_{\pi_{\pm}}^2+v_{\perp}^2(2n+1)|eB|+v_{\parallel}^2k_3^2}.\label{eq:Epc}
\end{align}
Here we use the strong-field form of charged pion's energy dispersion relation, to take the cyclotron motion into consideration. $m_{\pi_{\pm}}$ refers to the mass obtained by solving the pole equation with zero external momentum, and both $v_{\parallel}$ and $v_{\perp}$ are extracted from results in weak field limit. It should be mentioned that in strong field formulation, the velocity in transverse direction is ill-defined, since the momentum in corresponding direction has become separate Landau levels. To guarantee that in weak field limit \cref{eq:Epc} can reduced back to \cref{eq:mesonE}, we keep this coefficient in this work. In the following, the "propagating velocity" of $\pi_{\pm}$ at finite magnetic field should be understood as a ratio before Landau level. In \cref{appendixB}, we try to understand the definition of $v$ better through an demonstrative analysis on meson's two-point correlation function.

Now with the complete form of thermodynamic potential beyond mean field, we can get the new constituent quark mass from the corresponding gap equation with the feedback effect from meson,
\begin{align}\label{eq:gapbmf}
	m_q\left(\frac{1}{4G}+\frac{\partial\Omega_q}{\partial m_q^2}+\sum_M \frac{\partial\Omega_M}{\partial m_q^2}\right)=\frac{m_0}{4G}.
\end{align}
Comparing the mean field quark mass $m_{\text{MF}}$ from Eq.~(\ref{gap_mf}) with the newly obtained quark mass $m_q$ from \Cref{eq:gapbmf}, there is a mass difference which comes from the quantum fluctuations above the mean field. Following the procedure described in \cite{Zhuang:1994dw}, the meson contribution in thermodynamic potential can be expanded in power of $(m_q^2-m_{\text{MF}}^2)$,
\begin{eqnarray}
	\Omega_M=\sum_{n=0}^{\infty}\frac{1}{n!}\frac{\partial^n \Omega_M}{(\partial m_q^2)^n}\Big|_{m_q^2=m_{\text{MF}}^2}(m_q^2-m_{\text{MF}}^2)^n.
\end{eqnarray}

To simplify the calculation, we only consider the first two terms of the above series, where the leading order term vanishes and only the next-to-leading order term has none-zero contribution, then we get the practical form of gap equation, 
\begin{eqnarray}
	m_q\left(\frac{1}{4G}+\frac{\partial\Omega_q}{\partial m_q^2}+\sum_M \frac{\partial\Omega_M}{\partial m_q^2}\Big|_{m_q^2=m_{\text{MF}}^2}\right)=\frac{m_0}{4G}.
\end{eqnarray}

In the previous beyond-mean-field calculation with non-vanishing magnetic field, most works focus on the effect of neutral pions in the chiral limit, giving a conclusion that including a neutral pion with a physical propagating velocity may cause inverse magnetic catalysis effect. Those calculations stand because they are working in strong field limit, where charged pion gained a large mass from magnetic field and hence their contribution in the distribution function can be neglected. In this work, however, we work in weak field limit with $eB\sim m_{\pi}^2$. The mass disparity between neutral and charged pions is not significant enough, and the contributions from charged pions should not be overlooked. In the next section, we will
investigate how the neutral pion and charged pions  affect the chiral condensate and magnetic susceptibility.

\section{Numerical Results}
\label{sec:results}

In order to analyze the roles of neutral and charged pions in (inverse) magnetic catalysis and magnetic susceptibility, we consider the following four case: Case-0) Mean field approximation; Case-I) only neutral pion contribution; Case-II) only charged pions contribution; Case-III) both neutral and charged pions contribution, which are listed in TABLE I. We will also investigate the effect with and without finite pion propagating velocities separately.

Because of the contact interaction in the NJL model, the ultraviolet divergence cannot be eliminated through renormalization, and a proper regularization scheme is needed. In our work, we apply the gauge invariant Pauli-Villars regularization~\cite{Mao:2016fha}, which can guarantee the law of causality and effectively avoid the unphysical oscillation at finite magnetic field.
Other schemes, like proper-time regularization scheme and magnetic field independent regularization (MFIR)\cite{Coppola:2019uyr,Avancini:2021pmi}, which separates the vacuum and magnetic contribution, are also effective when dealing with system under a uniform magnetic field. By fitting the physical quantities, chiral condensate $\langle\bar{\psi}\psi\rangle=(-250\text{MeV})^3$, pion decay constant $f_{\pi}=93\text{MeV}$ in vacuum, we fix the current mass of light quarks $m_0=5\text{MeV}$, and obtain the parameter $\Lambda=1127\text{MeV}$. For different cases, the coupling constants are given in TABLE I.
\begin{table}[h]
    \centering
    \begin{tabular}{p{6em}|p{10em}|p{3em}}
    \hline
    \textbf{Case} & \textbf{Included mesons} & \textbf{$G\Lambda^2$} \\
    \hline
    0 & None &4.37 \\
    I & $\pi_0$ only & 4.81\\
    II & $\pi_{\pm}$ only & 5.36 \\
    III & $\pi_0,\ \pi_{\pm}$ & 6.05\\
    \hline
    \end{tabular}
    \caption{Coupling constants in different cases in and beyond mean field approximation.}
    \label{tab:my_label}
\end{table}

The full calculating procedure is as follows: 1) we first calculate the mean field gap equation Eq.~(\ref{gap_mf}) to get the mean field quark mass $m_{\text{MF}}$ before substituting it into the pole equation Eq.~(\ref{eq:poleeq}), where we can get the pole and screening masses of $\pi_0$ and $\pi_{\pm}$, and hence the corresponding propagating velocity in longitudinal and transverse directions. 2) Inserting the pion dispersion relation obtained above into the beyond-mean-field thermodynamic potential Eq.~(\ref{eq:bmf_potential}) and gap equation Eq.~(\ref{eq:gapbmf}), we can finally get the numerical result of order parameter $m_q$ and magnetic susceptibility. In this work, we employ this systematic step-by-step approach to go beyond the mean field approximation, and more self-consistent ways to go beyond mean field in absence of magnetic field are given in \cite{Huang:1999dz, Oertel:2000jp}.

In the case of explicit chiral symmetry breaking, the pseudo Nambu-Goldstone modes $\pi$ have a finite mass, making them propagating with a velocity lower than the speed of light at finite magnetic field and finite temperature. Due to the qualitative similarity in the behaviors of pion masses and propagating velocities across different cases, in \Cref{fig:mpeB} to \cref{fig:v_pi0} we will choose the case that exhibits the most conclusive results to highlight the distinctions arising from a relatively weak magnetic field.

\subsection{Meson properties}

\begin{figure}[htbp!]
		\includegraphics[width=\columnwidth]{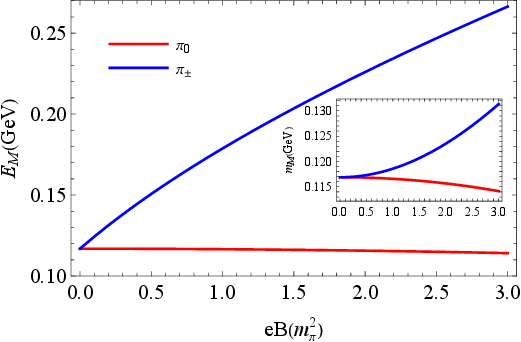}
		\caption{Neutral (Red solid line) and charged (Blue solid line) pion masses beyond mean field approximation in Case III as a function of magnetic field, with vanishing temperature.}\label{fig:mpeB}
\end{figure}

Here we start by examining the characteristics of neutral and charged pion masses at finite temperature and magnetic field. \Cref{fig:mpeB} demonstrates the magnetic field dependence of $\pi_0$ and $\pi_{\pm}$ masses at zero temperature. We use the weak field expansion and only calculate to $eB=3m_{\pi}^2$ with $m_{\pi}^2=0.0179\ \text{GeV}^2$. For $\pi_0$ mass, it decreases with the increase of the magnetic field, while for the ground state energy of $\pi_{\pm}$, which is given by $E_{\pi_{\pm}}=\sqrt{m_{\pi_{\pm}}(eB)^2+eB}$, it increases with the magnetic field, showing an opposite tendency. If we consider $\pi_0$ as a charge-neutral point-like particle, it should remain unaffected by external magnetic field, while in the NJL model where mesons are considered as composite particles made of quark anti-quark, $\pi_0$ mass changes with the changing of quark condensate. From Nambu-Goldstone theorem, since the explicit breaking of chiral symmetry will get more "explicit" with the restoration of chiral symmetry, the mass of the pseudo-Goldstone boson $\pi_0$ should generally decreases with magnetic field, which is consist with our numerical result. A uniform magnetic field breaks the isospin symmetry between $u$ and $d$ quarks, and $\pi_{\pm}$ gained mass from magnetic field, which is qualitatively consist with the point-like approximation (PLA) result, i.e.  $E_{\pi_{\pm}}^{\text{PLA}}(B)=\sqrt{m_{\pi_{\pm}}^2(B=0)+eB}$.

\begin{figure}[htbp!]
		\includegraphics[width=\columnwidth]{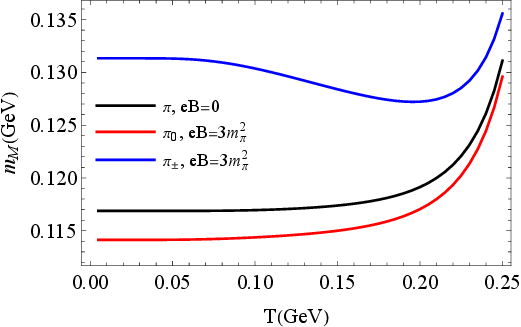}
		\caption{Pion mass beyond mean field approximation in Case III as a function of temperature with vanishing magnetic field.}\label{fig:mpT}
\end{figure}

In \cref{fig:mpT} we show the temperature-dependence of $m_{\pi_0,pole}$ and $m_{\pi_{\pm}}$ at $eB=0$ and $eB=3m_{\pi}^2$. At vanishing magnetic field, neutral and charged pions share the same mass. As the temperature gets higher, pion mass increases with the restoration of chiral symmetry, which is in agreement with the analysis of Nambu-Goldstone theorem.
At $eB=3m_{\pi}^2$, the pole masses of neutral and charged pions split. It should be noticed that at high temperature region, the system is thermalized and the splitting between $m_{\pi_0,pole}$ and $m_{\pi_{\pm}}$ becomes smaller.

\begin{figure}[htbp!]
		\includegraphics[width=\columnwidth]{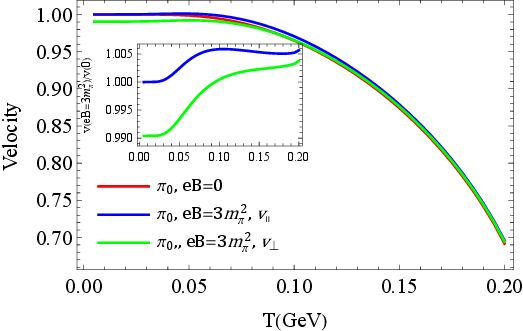}\\
		\includegraphics[width=\columnwidth]{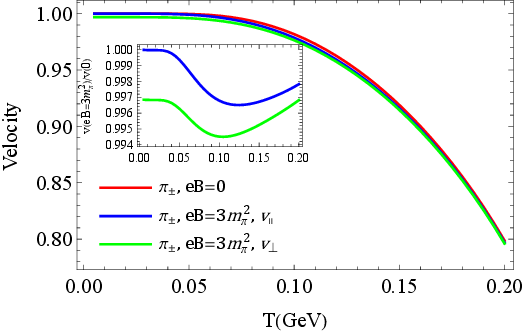}
		\caption{Propagating velocity (ratio) of $\pi_0$ (upper panel) and $\pi_{\pm}$ (lower panel) beyond mean field in Case I and Case II as a function of temperature at different magnetic field in the chiral symmetry breaking phase. The inlaid sub-figure shows the ratio of velocities at $eB=3m_{\pi}^2$ and $eB=0$ in corresponding directions.}\label{fig:v_pi0}
\end{figure}

\cref{fig:v_pi0} shows longitudinal and transverse velocities of both $\pi_0$ and $\pi_{\pm}$ as a function of the temperature at $eB=0$ and $eB=3m_{\pi}^2$. At vanishing magnetic field, both $\pi_0$ and $\pi_{\pm}$ exhibit isotropic behavior and their velocities decrease with the increase of temperature, since the temperature leads to a breakdown of  Lorentz symmetry in boost transformations. When a magnetic field is turned on, we can clearly observe the splitting between the longitudinal and transverse velocities for both $\pi_0$ and $\pi_{\pm}$. For transverse velocity, it gets lower than the vanishing magnetic field scenario, while for longitudinal velocity, it is generally the same as that in the vanishing magnetic field case at low temperature region. When the temperature gets higher, the splitting between the longitudinal and transverse velocities induced by magnetic field gets melted and the system tends to become isotropic again. This thermalization phenomenon is also observed in the strong field case, e.g.  \cite{Sheng:2020hge}. When the temperature increases, the velocity of neutral pion at $eB=3m_{\pi}^2$ gets larger than that at $eB=0$, since the neutral pion becomes lighter at higher magnetic field. The behavior of $\pi_{\pm}$ is basically the same as the neutral pion case, except that with the increase of magnetic field strength, the propagating velocity of $\pi_{\pm}$ tends to be smaller, given that the charged pion mass is increased by the magnetic field. 

It should be noted that in the chiral limit, the behavior of propagating velocities appears to exhibit the opposite trend, where the longitudinal velocity is always exactly the same as the speed of light, the transverse velocity gets lower as temperature increases. This discrepancy between transverse and longitudinal velocities was believed to lead to magnetic inhibition, which gives the IMC result in the chiral limit \cite{Fukushima:2012kc,Mao:2016fha,Mao:2017tcf}.

\subsection{Chiral condensate beyond mean field}

\begin{figure}
    \centering
    \includegraphics[width=\columnwidth]{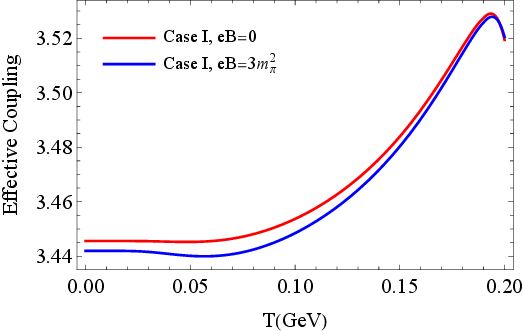}
    \includegraphics[width=\columnwidth]{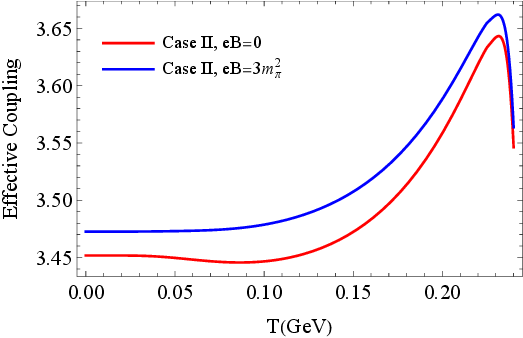}
    \includegraphics[width=\columnwidth]{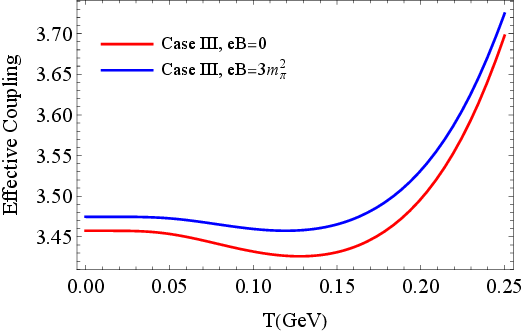}
    \caption{Effective coupling $\tilde{G}$ at $eB=0$ and $eB=3m_{\pi}^2$ as a function of temperature in three different cases, considering constant pion propagating velocities (ratios).}
    \label{fig:G_c}
\end{figure}

\begin{figure}
    \centering
    \includegraphics[width=\columnwidth]{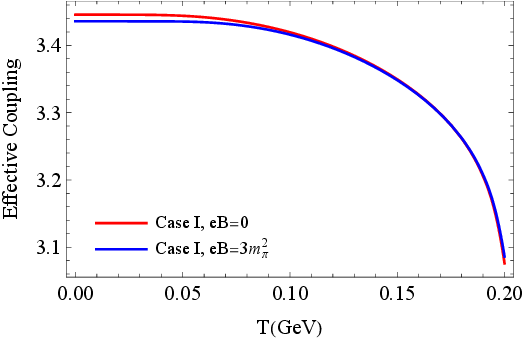}
    \includegraphics[width=\columnwidth]{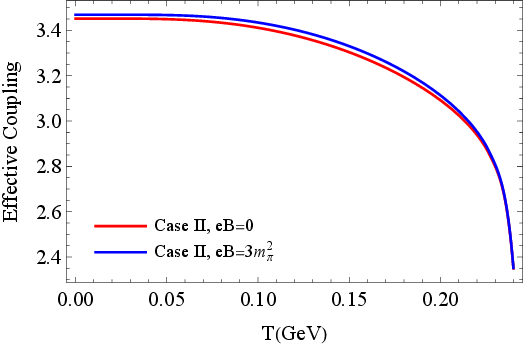}
    \includegraphics[width=\columnwidth]{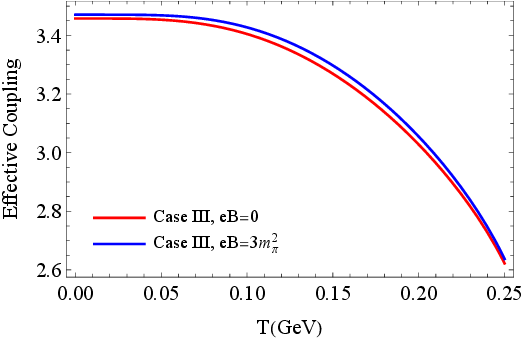}
    \caption{Effective coupling $\tilde{G}$ at $eB=0$ and $eB=3m_{\pi}^2$ as a function of temperature in three different cases, considering pion propagating velocities (ratios) dependent on temperature and magnetic field.}
    \label{fig:G_v}
\end{figure}

Now with the pole mass, screening masses and propagating velocities of $\pi_0$ and $\pi_{\pm}$ obtained in the mean-field approximation, we can insert them into the beyond-mean field thermodynamic potential and gap equation to get the quark mass in cases I,II and III. Before that, we can first calculate the effective coupling $\tilde{G}$ with 
\begin{align}
\label{eq:effectiveG}
\frac{1}{4\tilde{G}}\equiv\frac{1}{4G}+\sum_M \frac{\partial\Omega_M}{\partial m_q^2}\Big|_{m_q^2=m_{\text{MF}}^2}
\end{align}
to investigate the feedback effect from different pion to the system. To demonstrate the influence of magnetic field on effective coupling in the whole temperature region, in \cref{fig:G_c} and \cref{fig:G_v} we show $\tilde{G}(eB=0)$ and $\tilde{G}(eB=3m_{\pi}^2)$ with the feedback from different mesons, with and without the consideration of the effect of finite pion propagating velocities.

\cref{fig:G_c} shows the temperature dependence of effective coupling at different magnetic field strength with velocities $v=1$. We start with Case-I in the upper panel. The effective coupling generally increases until getting close to its Mott transition point, where a plunge is observed. As shown in \cref{eq:effectiveG}, the effective coupling is determined by the contribution from mesons. The mesonic energy dispersion relation is controlled by two different factors, i.e. the pole mass and propagating velocity. In \cref{fig:G_c}, velocities are set to be the speed of light, and as demonstrated in \cref{fig:mpT}, the pole mass of neutral pion is monotonically increasing with temperature, which leads to the increasing of mesonic energy and the decreasing of its contribution in the effective coupling. For the same reason, the effective coupling at higher magnetic field background always gets lower than that at vanishing magnetic field in Case I. For Case II and Case III in the middle and lower panels, Since we are examining the impact of charged pions, the effective coupling in these scenarios actually increases with the magnetic field. This is due to the fact that charged pions exhibit an opposing and more pronounced reaction to the magnetic field. This result suggests that neglecting the finite velocity effect may weaken magnetic catalysis in Case I, but strengthen it in Cases II and III. Since the inclusion of Schwinger phase does not bring qualitative change to the behavior of charged pion mass, according to \cite{Mao:2018dqe,GomezDumm:2023owj}, and the increment of $\pi_{\pm}$ masses still leads to the enhancement of effective coupling, we can conclude that the Schwinger phase will not affect our analysis.

\cref{fig:G_v} shows the temperature dependence of effective coupling at different magnetic field strength with finite velocities. Different from the result in \cref{fig:G_c}, the effective coupling in \cref{fig:G_v} monotonically decreases with the increase of temperature. This is because the drop in propagating velocities leads to a decrease in mesonic energy, which in turn increases the overall contribution from pions. We can also notice that, although the response of effective coupling to the magnetic field is qualitatively the same as in \cref{fig:G_c}, the magnitude is significantly weakened. In Case I, for example, the effect from magnetic field becomes indistinguishable at around temperature $T\sim 0.15\text{GeV}$. One can also find explanation from \cref{fig:mpT} and \cref{fig:v_pi0}, where the changes from propagating velocities counteracts the the effect of pole masses in energy dispersion relation.

\begin{figure}[htbp!]
		\includegraphics[width=\columnwidth]{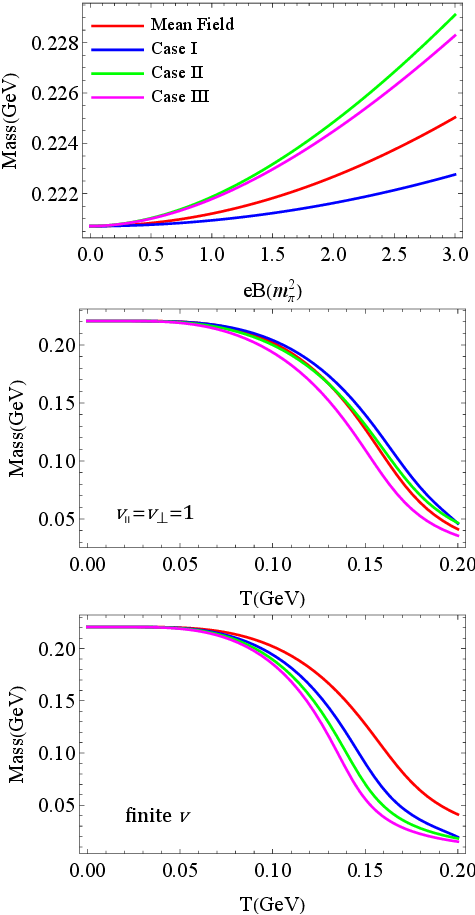}
		\caption{\label{fig:m}Effctive quark mass $m_q$ in four different cases, as a function of magnetic field with zero temperature (upper panel) and of temperature in vanishing magnetic field with finite and constant pion propagating velocities (ratios) (middle and lower panel), with $m_{\pi}^2\simeq0.0179\ \text{GeV}^2$.}
\end{figure}

In Fig.~\ref{fig:m}, we show the quark mass as a function of the magnetic field (up panel) and  the temperature with pion propagating velocity $v=1$ (middle panel) and finite $v$(below panel) for four cases listed in TABLE I. It is noticed from the upper panel that although the constituent quark mass increases with magnetic field in all cases, the increasing rate is lower in Case I with only neutral pion contribution compared to the mean-field case. The increasing rate of quark mass is significantly enhanced in both Case II and Case III in the case of considering the charged pion contribution. It is observed that in the middle figure where we consider the pionic velocity to be exactly the speed of light, the chiral symmetry restoration phase transition is catalysed in Case I and inverse-catalysed in Case III, while in Case II the behavior of quark condensate is similar to that of mean field case. However, in the lower panel, when the propagating velocities are dependent on the temperature and magnetic field strength, in all the beyond-mean-field cases, the chiral phase transition is inverse-catalysed, and the more pions are considered, the more inverse-catalytic the system becomes. By observing the results for different cases in \cref{fig:m}, it is noticed that the behaviors of these corrected quark masses agrees with the previous discussion in \cref{fig:G_c} and \cref{fig:G_v}.

\begin{figure}[htbp!]
		\includegraphics[width=\columnwidth]{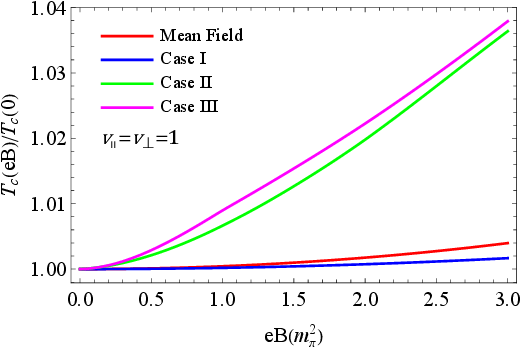}
		
		\includegraphics[width=\columnwidth]{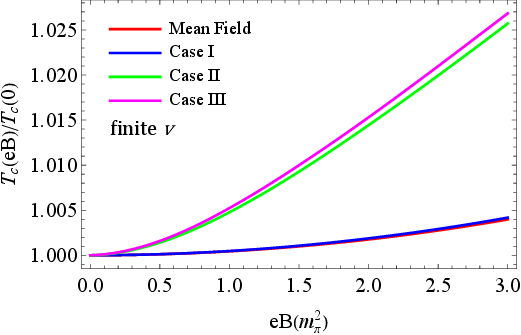}
		\caption{\label{fig:Tc_all}Critical temperature for chiral symmetry restoration phase transition in mean field approximation and three different cases beyond mean field as a function of magnetic field, normalized by vanishing magnetic field critical temperature, considering both constant pion propagating velocities (ratios) (upper panel) and finite pion propagating velocities (ratios) (lower panel).}
\end{figure}

To identify the magnetic field effect on the effective quark mass at high temperature region, in \Cref{fig:Tc_all} we show the re-scaled $T_c-eB$ phase diagram in case-I,II, and III with and without the consideration of finite pionic velocities, compared with the mean field result. In the upper panel with constant pion propagating velocities, the critical temperature in different cases is as follows: Case 0 (mean field): $T_c(0)=157.2 \text{MeV}$, Case I: $T_c(0)=163.5 \text{MeV}$, Case II: $T_c(0)=159.8 \text{MeV}$, Case III: $T_c(0)=151.6 \text{MeV}$; and in the lower panel with finite pion propagating velocities Case 0: $T_c(0)=157.2 \text{MeV}$, Case I: $T_c(0)=145.3 \text{MeV}$, Case II: $T_c(0)=139.5 \text{MeV}$, Case III: $T_c(0)=135.6 \text{MeV}$. The critical temperature $T_c$ is determined by the fastest drop of effective quark mass $\partial^2 m_q(T_c,eB)/\partial T^2=0$. It is noticed that in Case I where we only include the feedback effect from $\pi_0$, the increasing rate of critical temperature is weakened but the system is still magnetic catalytic. However, if we include the charged pions contribution, the increasing rate of critical temperature get enhanced. Once we consider the finite velocity effect, both the enhancement from charged pions and the weakening from neutral pion to magnetic catalysis are suppressed, just as discussed in \cref{fig:G_v}.

In the previous study of beyond-mean field calculation \cite{Fukushima:2012kc,Mao:2016fha,Mao:2017tcf}, IMC is achieved when a neutral pion with finite velocity is included into the system, due to the magnetic inhibition effect where the splitting between the longitudinal and transverse velocities get larger as the magnetic field and temperature are increasing. However, our calculation in this work gives an opposite result: Whether or not the influence of finite velocities is taken into account, the system remains magnetic catalytic.  The main difference between the previous study and this work is that, the previous study works in the chiral limit and in this work the chiral symmetry is explicitly broken. For transverse velocity, both previous and our present result share the same form, but for longitudinal velocity, the chiral limit version is set to be the speed of light since it is a massless particle, unlike our cases, where the splitting in velocities induced by magnetic field will be thermalized once the temperature goes near the critical temperature.

\subsection{Magnetic susceptibility}

Recently the lattice QCD calculation showed that the magnetized QCD matter exhibits diamagnetism at low temperature and paramagnetism at high temperature, so in this work we also calculate the magnetic susceptibility in the NJL model beyond mean field.

The magnetic susceptibility is defined by
\begin{eqnarray}
\bar{\chi}(T)=-\frac{\partial^2 \Omega(T,eB)}{\partial eB^2}\Big|_{eB=0}.
\end{eqnarray}
In lattice calculation, the renormalization scale choice fixes $\bar{\chi}(0)=0$ so that the divergence in vacuum magnetic susceptibility can be eliminated. Mimicking their procedure, we define a new magnetic susceptibility to ensure the relation $\chi(0)=0$, where
\begin{eqnarray}
\chi(T)=\bar{\chi}(T)-\bar{\chi}(0)=\frac{\partial^2\Delta(T,eB)}{\partial eB^2}\Big|_{eB=0},
\end{eqnarray}
with
\begin{align}
    \Delta(T,eB)=\Omega(0,eB)-\Omega(T,eB).
\end{align}

\begin{figure}[htbp!]
		\includegraphics[width=\columnwidth]{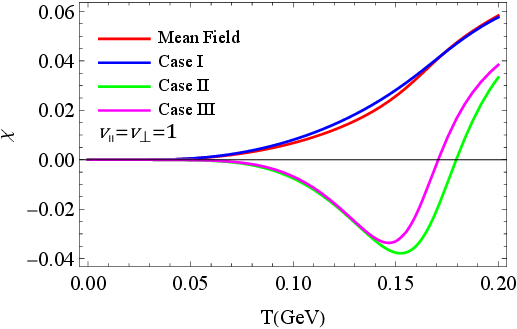}
		
		\includegraphics[width=\columnwidth]{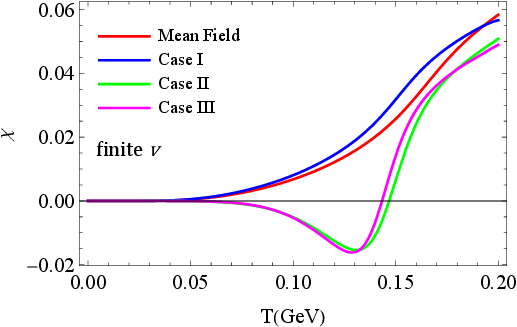}
		\caption{Magnetic susceptibility in mean field approximation (Red solid lines), beyond mean field with the feedback from $\pi_0$ (Blue solid lines), $\pi_{\pm}$ (Green solid lines) and $\pi_0$, $\pi_{\pm}$ (Cyan solid lines) as a function of temperature with vanishing magnetic field. Propagating velocity for pions are set to be speed of light in upper panel, and finite velocities effect is considered in lower panel.}\label{fig:chi}
\end{figure}

We numerically calculate the magnetic susceptibility in the vicinity of $B=0$, and show it in \Cref{fig:chi}. In mean field approximation, the magnetic susceptibility is always positive and increases with temperature. In the Case I where $\pi_0$ is included, there is no qualitative change in the behavior of magnetic susceptibility, but instead quantitatively it is enhanced by the feedback effect of $\pi_0$. However, if we include the effect of charged pions as well like in Case II and Case III, the magnetic susceptibility is negative at low temperature and becomes positive at high temperature region, showing qualitative agreement with the lattice result.

In the mean field approximation, the system only consists of quark, a charged particle with non-zero spin, which means its magnetic susceptibility is subject to two different mechanisms, namely Pauli paramagnetism and Landau diamagnetism \cite{Landau:dia1,Landau:dia2}. The former originate from the spin distribution in the presence of magnetic field, and the latter is a quantum effect that is related to the cyclotron motion of charged particles. Normally the Pauli paramatnetism is stronger than Landau diamagnetism, making the total magnetic susceptibility positive in the mean field approximation. However, the pseudo-scalar particles $\pi_{\pm}$ are spin-0 and only subject to Landau diamagnetism, giving a negative magnetic susceptibility result. When we go beyond mean field approximation, we actually consider a system composed of quarks and mesons. At low temperature, the pion meson contribution is dominant and the total magnetic susceptibility is negative, while at high temperature, the dominance of quark contribution makes the system going back to paramagnetism.
\section{Conclusion}
\label{conclusion}
In this work, we investigated the magnetic field effect on the phase transition of chiral symmetry restoration in the framework of $SU(2)$ NJL model beyond mean field approximation. We consider three cases, by separately including the effect of $\pi_0$ (Case I), $\pi_{\pm}$ (Case II), and all three pions altogether (Case III). Since sigma meson is much heavier than pions and it is not directly influenced by the magnetic field, we neglect its contribution in the calculation. To identify the effect of finite pionic propagating velocities, we consider two situations, one setting pion velocity to be exactly speed of light, and another with finite velocities determined by \cref{eq:v}. Unlike the previous IMC results in the chiral limit due to the magnetic inhibition effect where the splitting between the longitudinal and transverse velocities of pion suppressed the critical temperature, in this work in chiral symmetry explicitly broken case, the system is always magnetic catalytic. In the discussion of effective coupling, we notice that in absence of magnetic field, both neutral and charged pions share the same contribution to the system. However, when a magnetic field is introduced, due to the opposite response of charged and neutral pion masses to it, the feedback from $\pi_0$ weakens the MC but that from $\pi_{\pm}$ instead enhances it. Once we consider the effect from finite velocities, it will counter the effect from the changes in pionic pole mass. When we go to finite temperature region, all the effect above will be thermalized, resulting that IMC cannot be observed in any of the above scenarios. One problem in the calculation is in \cref{eq:Epc} where we consider a propagating velocity of $\pi_{\pm}$ from weak field expansion in the energy dispersion relation in strong-field formulation, since the transverse velocity here is ill-defined with momentum in corresponding direction becoming quantized Landau levels. In our future work, we plan to do the calculation in a more self-consistent way, e.g. rewrite \cref{eq:Epc} in weak-field expansion manner.

We also calculate the magnetic susceptibility in these three scenarios. In mean field approximation and only including $\pi_0$ scenario, the magnetic susceptibility is always positive and increases with temperature. However, if we include the effect of charged pions, the magnetic susceptibility becomes negative at low temperature and positive at high temperature region, showing qualitative agreement with the lattice result. A system composed of quarks is governed by both Landau diamagnetism and Pauli paramagnetism. The former originates from the spin distribution in the presence of magnetic field, and the latter is a quantum effect which relates to the cyclotron motion of charged particles. Normally the Pauli paramagnetism is stronger than Landau diamagnetism, making the total magnetic susceptibility positive in the mean field approximation. For charged pions which is spin zero, its magnetic susceptibility is only controlled by Landau diamagnetism. In beyond mean field approximation, the system is composed of quarks and mesons, and at low temperature, the meson contribution is dominant and the total magnetic susceptibility is negative, while at high temperature, the dominance of quark contribution makes the system going back to be paramagnetic.

\hskip 2cm
\begin{acknowledgments}

In this work, J. M., R. W., M. H. and K. X. are supported in part by the National Natural Science Foundation of China (NSFC) Grant Nos: 12235016, 12221005, 12147150 and the Strategic Priority Research Program of Chinese Academy of Sciences under Grant No XDB34030000 and and Fundamental Research Funds for the Central Universities, and S. M. is supported by the NSFC Grant No. 12275204 and Fundamental Research Funds for the Central Universities. 

\end{acknowledgments}

\appendix

\begin{widetext}
\section{The sub-leading order of polarization functions}
\label{appendixA}
In this part we will derive the specific form of polarization function in weak field expansion. We start with the equation of $\Pi^{00}(q^2)$. For the pole mass, with the static condition, the $B=0$ contribution of the polarization function can be written as
\begin{align}
   \Pi^{00}(q_0^2,0)=-I_1^{(0)}(m_q^2)+q_0^2I_2^{(0)}(m_q^2,q_0^2,0),\nonumber\\ 
\end{align}
and $I_2^{(0)}(m_q^2,q_0^2,0)$ is given by
\begin{eqnarray}
I_2^{(0)}(m_q^2,q_0^2,0)=2\int\!\!\frac{d^3 k}{(2\pi)^3}\frac{\tanh\left(\frac{E_q}{2T}\right)}{E_q(4E_q^2-q_0^2)}.
\end{eqnarray}
Besides, the formula of $\Pi^{00}(0,q_i^2)$ for screening mass reads
\begin{align}
	\Pi&^{00}(0,q_i^2)=-\frac{T}{4\pi}\sum_{l=-\infty}^{\infty}\frac{1}{\sqrt{-q_i^2}}\bigg[\text{arctan}\left(\frac{2m_l}{q_i}\right)-\text{arctan} \left(\frac{2m_l-\sqrt{-q_i^2}}{q_i}\right)+\text{arctan}\left(\frac{2q_i}{2m_l-\sqrt{-q_i^2}}\right)\bigg],
\end{align}
with $m_l^2=m_q^2+\omega_l^2$ and Matsubara frequency $\omega_l=(2l+1)\pi T$.
Here we start with $\Pi^{20}$. Substituting \Cref{propagator_wf} into Eq.~(\ref{piij}), we have

\begin{align}
\Pi^{20}(q^2)&=2i\int \frac{d^4k}{(2\pi)^4} \text{Tr}_{\{s\}}\!\left[\gamma_5 \frac{k_{\perp}^2 (m+k\!\!\!/_{\parallel})+k\!\!\!/_{\perp} (m^2-k_{\parallel}^2)}{(k^2-m^2)^4}\gamma_5 \frac{m+k\!\!\!/-q\!\!\!/}{(k-q)^2-m^2}\right]\nonumber\\
&=8i\int \frac{d^4k}{(2\pi)^4}\frac{k_{\perp}^2[m^2-k_{\parallel}\!\!\cdot(k-q)_{\parallel}]-k_{\perp}\!\!\cdot(k-q)_{\perp}(m^2-k_{\parallel}^2)}{(k^2-m^2)^4[(k-q)^2-m^2]}.
\end{align}
To calculate the pole and screening mass in different directions in a form that is as simplified as possible, we should consider these three situations separately. For pole mass, we set the external momentum $q_{\perp}=q_3=0$ and induce a replacement $\int\frac{d^4k}{(2\pi)^4}\rightarrow iT\sum_{n=-\infty}^{\infty}\int\frac{d^3k}{(2\pi)^3},\ \ k_0\rightarrow i\omega_n$ in finite temperature case, to have\\
\begin{align}
\Pi^{20}(q_0^2,q^2_{\perp}=q^2_3=0)=-8T\sum_{n=-\infty}^{\infty}\int \frac{d^3k}{(2\pi)^3}\frac{ (i\omega_n) q_0k_{\perp}^2}{[(i\omega_n)^2-E_q^2]^4\{(i\omega_n-q_0)^2-E_q^2\}}.
\end{align}
One can choose to either do the summation of Matsubara frequencies numerically, or use the package \verb+MatsubaraSum+ to do the calculation.

For screening mass, according to \cite{Sheng:2020hge} the Matsubara summation should be done after the integration of internal momentum. Follow the similar procedure as above, we can set $q_0=0$ and get
\begin{align}\label{eq:Pi20m}
\Pi^{20}(q_0^2=0,q_{\perp}^2,q_3^2)=8i\int \frac{d^4k}{(2\pi)^4}\frac{k_{\perp}^2[m^2+k_3\!\!\cdot(k-q)_3]-k_{\perp}\!\!\cdot(k-q)_{\perp}(m^2+k_3^2)-k_{\perp}q_{\perp}k_0^2}{(k^2-m^2)^4[(k-q)^2-m^2]}.
\end{align}
Above we introduce an integration over Feynman parameters, which is given by

\begin{align}\label{eq:FeynmanP}
\frac{1}{(k^2-m^2)^a[(k-q)^2-m^2]^b}&=\int_0^1 dx\frac{x^{a-1}(1-x)^{b-1}}{\{x(k^2-m^2)+(1-x)[(k-q)^2-m^2]\}^{a+b}}\times\frac{\Gamma(a+b)}{\Gamma(a)\Gamma(b)}\nonumber\\
&=\int_0^1 dx\frac{x^{a-1}(1-x)^{b-1}}{(l^2-\Delta)^{a+b}}\times\frac{\Gamma(a+b)}{\Gamma(a)\Gamma(b)},
\end{align}

where
\begin{align}
    \Delta&=m^2-x(1-x)q^2,\nonumber\\
    l&=k-(1-x)q.
\end{align}

Substituting Eq.~(\ref{eq:FeynmanP}) into Eq.~(\ref{eq:Pi20m}), setting $k=l+(1-x)q$ and discarding all the odd powers of $l$, we obtain
\begin{align}
    \Pi^{20}(q_0^2=0,q_{\perp}^2,q_3^2)=-32T\sum_{n=-\infty}^{\infty}\int \frac{d^3l}{(2\pi)^3}\int_0^1 dx \frac{x^3(1-x)(-l_{\perp}^2q_3^2+m^2q_{\perp}^2+l_3^2q_{\perp}^2-(i\omega_n)^2q_{\perp}^2)}{((i\omega_n)^2-l_3^2-l_{\perp}^2-\Delta)^5}.
\end{align}

Separately set $q_{\perp}^2=0$ and $q_3^2=0$ and performing the integration in $l$, we can get the $\Pi^{20}$ in longitudinal and transverse directions, accordingly,

\begin{align}
    \Pi^{20}(q_0^2=q_{\perp}^2=0,q_3^2)&=-\frac{T}{8\pi}\sum_{n=-\infty}^{\infty}\int_0^1 dx \frac{x^3(1-x)q_3^2}{(\Delta-(i\omega_n)^2)^{5/2}},\nonumber\\
    \Pi^{20}(q_0^2=q_3^2=0,q_{\perp}^2)&=\frac{T}{16\pi}\sum_{n=-\infty}^{\infty}\int_0^1 dx \ x^3(1-x)\left\{\frac{q_{\perp}^2}{(\Delta-(i\omega_n)^2)^{5/2}}+\frac{5(m^2-(i\omega_n)^2)q_{\perp}^2}{(\Delta-(i\omega_n)^2)^{7/2}}\right\}.
\end{align}

Following the similar procedure, we can get the expression for $\Pi^{11}$:

\begin{align}
    \Pi^{11}(q_{\perp}^2=q_3^2=0,q_0^2)&=4T\sum_{n=-\infty}^{\infty}\int \frac{d^3k}{(2\pi)^3}\frac{m^2+k_3^2-(i\omega_n)(i\omega_n-q_0)}{[(i\omega_n)^2-E_q^2]^2\{(i\omega_n-q_0)^2-E_q^2\}^2},\nonumber\\
    \Pi^{11}(q_0^2=q_{\perp}^2=0,q_3^2)&=\frac{T}{8\pi}\sum_{n=-\infty}^{\infty}\int_0^1 dx\  x(1-x)\left\{\frac{1}{(\Delta-(i\omega_n)^2)^{3/2}}+\frac{3[m^2-x(1-x)q_3^2-(i\omega_n)^2]}{(\Delta-(i\omega_n)^2)^{5/2}}\right\},\nonumber\\
    \Pi^{11}(q_0^2=q_3^2=0,q_{\perp}^2)&=\frac{T}{8\pi}\sum_{n=-\infty}^{\infty}\int_0^1 dx \ x(1-x)\left\{\frac{1}{(\Delta-(i\omega_n)^2)^{3/2}}+\frac{3[m^2-(i\omega_n)^2]}{(\Delta-(i\omega_n)^2)^{5/2}}\right\}.
\end{align}
For vanishing temperature case, the sub-leading order $\Pi^{02}$ and $\Pi^{11}$ have the form 

\begin{eqnarray}
	&&\!\!\!\Pi^{20}(q_{\parallel}^2,q_{\perp}^2)(T=0)=\frac{1}{6\pi^2}\int_{0}^{1}\!\!dx \frac{x^3(1-x)}{\Delta^3}\left(2m_q^2q_{\perp}^2+(q_{\perp}^2+q_{\parallel}^2)\Delta\right),\ \ \ \ \ \ \nonumber\\
	&&\!\!\!\Pi^{11}(q_{\parallel}^2,q_{\perp}^2)(T=0)=\frac{1}{4\pi^2}\int_{0}^{1}\!\!dx \frac{x(1-x)}{\Delta^2}\left(m_q^2+x(1-x)q_{\parallel}^2+\Delta\right),
\end{eqnarray}
The above expression works well in calculating both pole and screening masses for pions.

\section{Inclusion of velocity (ratio)}
\label{appendixB}
In this section we try to understand the propagating velocity (or ratio) given in \cref{eq:NMeson} and \cref{eq:Epc} better with a demonstrative analysis on the propagator of a meson particle.

Firstly, let's consider the 2-point correlation function (the inverse of propagator) for a meson in Euclidean space without magnetic fields
\begin{align}
    \Gamma^{(2)}=Z_0 p_0^2+Z_{\vec{p}} \vec{p}^2 +m_0^2.
\end{align}
Here $Z_0$ and $Z_{\vec{p}}$ denote the temporal and spatial wave function renormalizations, and $m_0$ are bare meson mass. The pole mass and screening masses are defined as \cite{Helmboldt:2014iya}
\begin{align}
    \Gamma^{(2)}(p_0= i m_{pole},\vec{p}=0)&=0\,, \\
    \Gamma^{(2)}(p_0= 0,\vec{p}^2=- m_{\text{scr}}^2)&=0\,.
\end{align}
By solving the equations above, we get
\begin{align}
    Z_0=\frac{m_0^2}{m_{\text{pole}}^2}\,, \quad\quad Z_{\vec{p}}=\frac{m_0^2}{m_{\text{scr}}^2}\,,
\end{align}
the ratio of the temporal and spatial wave function renormalizations is
\begin{align}
    v\equiv \sqrt{\frac{Z_{\vec{p}}}{Z_0}}=\frac{m_{pole}}{m_{\text{scr}}},
\end{align}
That's Eq.(30) in our paper. The energy dispersion relation is
\begin{align}
    E=\sqrt{m_{pole}^2+v^2 \vec{p}^2}.
\end{align}

Then, we consider the translational invariant part of the propagator for a charged scalar meson under magnetic fields in the Schwinger scheme:
\begin{align}
    G(p)=\int_0^\infty \frac{i ds}{\cos(qBs)} \exp\left(-is(Z_0 p_0^2 + Z_{3} p_3^2 + Z_{\perp} p_\perp^2 \frac{\tan(qBs)}{qBs} +m_0^2)\right)\,,
\end{align}
we make the change of variable $s\rightarrow -i \tau/(qB)$
\begin{align}
    G(p)=\frac{1}{qB}\int_0^\infty \frac{d \tau}{\cos(-i \tau)} \exp\left(-\frac{\tau}{qB}(Z_0 p_0^2 + Z_{3} p_3^2 + Z_{\perp} p_\perp^2 \frac{\tan(-i \tau)}{-i \tau} +m_0^2)\right).
\end{align}
By using
\begin{align}
    \cos(i \tau)=\frac{e^\tau+e^{-\tau}}{2}\,, \quad \quad i\tan(-i \tau)=\frac{e^\tau-e^{-\tau}}{e^\tau+e^{-\tau}}\,,
\end{align}
and the generating function of the Laguerre polynomials
\begin{align}
\frac{\exp(-xz/(1-z))}{1-z}=\sum_{l=0}^\infty L_l (x) z^l\,,
\end{align}
we arrive
\begin{align}
    G(p)&=\frac{2}{qB}\sum_{l=0}^\infty L_l\bigg(\frac{Z_\perp p_\perp^2}{qB}\bigg) (-1)^l e^{-\frac{Z_\perp p_\perp^2}{qB}} \int_0^\infty d\tau \exp\left(-\frac{\tau}{qB} (Z_0 p_0^2+Z_3 p_3^2+(2l+1)qB +m_0^2)\right) \nonumber \\
    &=\frac{2}{qB}\sum_{l=0}^\infty L_l\bigg(\frac{Z_\perp p_\perp^2}{qB}\bigg) (-1)^l e^{-\frac{Z_\perp p_\perp^2}{qB}} \frac{qB}{Z_0 p_0^2+Z_3 p_3^2+(2l+1)qB +m_0^2}.
\end{align}
The energy dispersion relation reads
\begin{align}
    E=\sqrt{\frac{m_0^2+Z_3 p_3^2+(2l+1)qB}{Z_0}},
\end{align}
where $m_0/\sqrt{Z_0}$ is $m_{\pi_{\pm}}$ in this version of our article. Here, all the Landau levels comes with a coefficient $1/Z_0$. 

It should be mentioned that above is only a demonstrative analysis. 
In NJL model, since pions are composite particles, the value of $Z_0$ cannot be determined, and 
we use the $v_\perp^2$ as in \cref{eq:Epc} for the coefficients before the Landau levels instead.

\end{widetext}

\bibliography{apssamp}

\end{document}